# Demonstration of the existence of biased transportation in the asymmetric nanoscale systems induced by thermal noise from a simple mathematics perspective

Haiping Fang*, Rongzheng Wan

*Division of Interfacial Water, Shanghai Institute of Applied Physics, Chinese Academy of Sciences, 201800 Shanghai, China*
*fanghaiping@sinap.ac.cn*

**The existence of net flux across the nanochannels under the asymmetrical static electric fields obtained at room temperature from molecular dynamics (MD) simulations [1, 2] has been extensively queried on whether the second law of the thermodynamics holds, and whether the observations result from the numerical artifacts. Here we use a simple mathematics to demonstrate the existence of biased-directional transportation (net flux across the channel) in those asymmetrical nanoscale systems at room temperature, at least in the MD simulations. We find that the key to the existence of biased-directional transportation is that the thermal noise is not white anymore at nanoscale; it includes the color noise with an autocorrelation time length (for MD simulation, this length is of ~10 ps [3]).**

We only consider the simulations and leave the physics apart. In the numerical simulations, we usually make the quantities of the system non-dimensional. This dimensionless system can return the nanoscale system by the original scaling parameters, as well as describe a mesoscopic system by new scaling parameters.

For the mesoscopic system, we can keep the asymmetry of the system without any input of energy, and importantly, the biased-directional transportation can be induced by the fluctuations with a large enough autocorrelation time length, according to the ratchet theory at the mescoscopy [4-6]. We all agree that this biased-directional transportation results from the energy provided by the fluctuations with an autocorrelation time length that is large enough.

It should be noted that the numerical results for this mesoscopic system are consistent with the numerical results for the nanoscale system, only with the difference in units, since they are both from the same dimensionless system. For example, the nanoscale pumping system in Ref. 1 (The length of the system is only 1 nm and the thermal noise includes the color noise with an autocorrelation time length of ~ 10 ps) can be rescaled to a mesoscopic system of 100 μm in length with fluctuations including the color noise with an autocorrelation time length of 0.1 μs.

Since we can accept the existence of the biased-directional transportation in the mesoscopic system rescaled from the nanoscale system, the existence of the biased-directional transportation in the original nanoscale system can be expected.

Thus, from the same dimensionless model, we can obtain two real systems
- Mesoscopic system: Fixing of the mesoscopic structures does not require any external energy input; the energy input is provided by the fluctuations with mesoscopic autocorrelation time lengths. Biased-directional transportation results from the energy input provided by the fluctuations.
- Nanoscale system: Fixing of the nanoscale structures in an asymmetric state (may) requires external energy input; (According to traditional theory,) no energy input is provided by the thermal noises with nanoscale autocorrelation time lengths. The existence of the biased-directional transportation in this asymmetrical nanoscale system at the room temperature is reliable, but the energy input for the biased-directional transportation is still not clear.

For this nanoscale system, new questions will be raised about how the energy for biased-directional transportation comes from and what the route is for the energy transfer? At nanoscale, we do not expect the energy provided by the thermal fluctuation even it is actually a color noise with an autocorrelation time length that is large enough (This is from a traditional theory from mesoscopic and macroscopic ratchet theory. I do not know whether it is correct at nanoscale but we first assume it is true.), but the position constraint of asymmetrical structure is not easy since the thermal fluctuation becomes important which will drift them from their initial positions. We believe that the constraint requires external energy and the question arises that how the energy used to constrain the system transfers to the drive of the biased-directional transportation.

Finally, there is one point to be noted. The mesoscopic system rescaled from a nanoscale system may not describe a real system. However, from the ratchet theory, there is biased-directional transportation in this mesoscopic system described by equations and fluctuations mathematically.

The idea can be clearer from the mathematics below by taking the pumping system in Ref. 1 as an example.

In the MD simulation, the motion of a water molecule inside the nanotube can be written in the form of

$$m\ddot{x}_i(t) = -V'(x_i(t)) + f(x_i,t)$$ (1)

here, $m$ is the mass of the water molecule, $x_i$ is the coordinate of an water molecule, $V(x_i)$ is the potential between this water molecule and the nanochannel including the charges, when the charges are fixing, $V(x_i)$ does not change with time $t$, $f(x_i, t)$ has the form of

$$f(x_i,t) = \sum_j \frac{dV_{LJ}(r_{ij})}{dr_{ij}} \frac{\mathbf{r}_{ij}}{r_{ij}} + \sum_j \frac{dV_C(r_{ij})}{dr_{ij}} \frac{\mathbf{r}_{ij}}{r_{ij}} + \xi(t)$$ (2)

$f(x_i, t)$ includes the total LJ interactions and Coulomb interactions between this water molecule and the surrounding water molecules, $\xi(t)$ is a stochastic part introduced by the T coupling method.

We introduce dimensionless units, (1) can be written in the form of

$$\hat{m}\ddot{\hat{x}}_i(\hat{t}) = -\hat{V}'(\hat{x}_i(\hat{t})) + \hat{f}(\hat{x}_i,\hat{t})$$ (3)

$\hat{m}$ is the dimensionless mass, $\hat{x}$ is the dimensionless coordinate, $\hat{V}(\hat{x})$ is the dimensionless asymmetric potential, $\hat{t}$ is the dimensionless time and $\hat{f}(\hat{t})$ is the dimensionless force. The relationship between the real units and the dimensionless units is

$$\Delta V = \frac{mL^2}{t^2} \frac{\Delta \hat{V} \hat{t}^2}{\hat{L}^2 \hat{m}}$$ (4)

For the nanoscale MD system in Ref. 1, we have a mass unit of 1.67E-27 kg, length unit of 1 nm, time unit of 1 ps, thus we have an energy unit of 1.67E-21 J($\approx$0.4 $k_B$T, when T=300 K)

From equation (3), we can get the dimensionless probability density $\hat{P}(\hat{x},\hat{t})$, and the dimensionless probability current $\hat{J}(\hat{x},\hat{t})$ defined as

$$\hat{P}(\hat{x},\hat{t}) = \langle \delta(\hat{x}-\hat{x}(\hat{t})) \rangle$$ (5)

$$\hat{J}(\hat{x},\hat{t}) = \langle \dot{\hat{x}}(\hat{t})\delta(\hat{x}-\hat{x}(\hat{t})) \rangle$$ (6)

This dimensionless equation (3) can also describe a mesoscopic system, in which the real mass unit can be 1.67E-3 μg, the time unit can be 0.1 μs, and the length unit can be 100 μm. For such a mesoscopic system, it is acceptable that an autocorrelated fluctuation $\hat{f}(\hat{t})$ with an asymmetric potential $\hat{V}(\hat{x})$ results in a non-zero current $\hat{J}(\hat{x},\hat{t})$, and the energy comes from the fluctuation.

Then we come back to the nanoscale system, which can be described by the same dimensionless equation (3), only with different units. It is clear that the existence of the net flux from the equation (3) does not change with such rescaling. Since the existence of the unidirectional flux in the mesoscopic system rescaled from the nanoscale system is acceptable, the existence of the unidirectional flux in this nanoscale system from MD simulation is reasonable.